

\input harvmac

\def\etal{{\it et al.}}
\def\b{{\rm b}}
\def\s{{\rm s}}
\def\ol{\overline}
\def\e{{\rm e}}
\def\O{{\cal O}}
\def\A{{\cal A}}
\def\H{{\cal H}}
\def\d{{\rm d}}
\def\im{{\rm i}}
\def\q{{\rm q}}
\def\vslash{v\hskip-0.5em /}
\def\Aslash{A\hskip-0.5em /}
\def\mev{\;{\rm MeV}}
\def\gev{\;{\rm GeV}}
\def\OMIT#1{}
\def\frac#1#2{{#1\over#2}}
\def\eps{{\epsilon}}
\def\pintegral{\int{\d^{4-\eps}\,p\over(2\pi)^{4-\eps}}}
\def\gpref{{g^2\over16\pi^2f^2}}
\def\pref{{1\over16\pi^2f^2}}

\hbadness=10000

\Title{\vbox{\hbox{SLAC--PUB--6237}\hbox{SSCL--Preprint--422}
\hbox{hep-ph@xxx/9306310}}}
{$\ol B\to \ol K\e^+\e^-$ in Chiral Perturbation Theory}

\centerline{Adam F. Falk}
\bigskip{\it\centerline{Stanford Linear Accelerator Center}}
{\it\centerline{Stanford University, Stanford, California 94309}}
\bigskip\bigskip
\centerline{Benjam\'\i n Grinstein}
\bigskip{\it\centerline{Superconducting Supercollider Laboratory}}
{\it\centerline{2550 Beckleymeade Avenue, Dallas, Texas 75237}}

\vskip .3in

The combined use of chiral $SU(3)$ and heavy quark symmetries allows
one to relate the hadronic form factors for the decay $\ol B\to
\ol K\e^+\e^-$ to those for $\ol B\to\pi\e^-\ol\nu$. We investigate
departures from the symmetry limit which arise from chiral symmetry
breaking. The analysis uses chiral perturbation theory and the heavy
quark limit to compute the relevant hadronic matrix elements. We
estimate the size of $SU(3)$ corrections by computing, at one loop
order, the leading nonanalytic dependence on the light quark masses.
The calculation is trustworthy only in the portion of the Dalitz plot
in which the momentum of the kaon or pion is small. We find the
corrections to be~$\sim40\%$.

\Date{June 1993}
\vfil\eject

Flavor changing neutral transitions are suppressed in the standard
model of electroweak interactions, because they do not occur at tree
level, and at one loop because of the GIM mechanism. In $s\to d$
transitions GIM cancellations are very effective for diagrams involving
virtual $u$ and $c$ quarks, while  virtual $t$ diagrams are doubly
Cabibbo suppressed relative to the transitions mediated by $u$ and $c$.
By contrast, in $b\to s$ transitions it is those diagrams involving
virtual $u$ quarks which are doubly Cabibbo suppressed; then the GIM
cancellation is rather ineffective, as it involves the very heavy $t$
quark against the light (by comparison) $c$ quark.

Hence processes involving $b\to s$ flavor change are interesting
because, although rare, they are well within experimental reach. In
fact, the first measurement of such a process was recently reported by
the CLEO collaboration, who observe the process $B \to K^*\gamma$ with
a branching fraction of $(4.5\pm1.9\pm0.9)\times10^{-5}$ \ref\cleo{R.
Ammar \etal, CLNS 93/1212 and CLEO 93-06 (1993)}. More importantly,
however, processes involving $b\to s$ flavor change are interesting
because, being rare, they are a quite sensitive probe of departures
from standard expectations. There exist several studies of the effect
of extensions to the standard model on the rates for this class of
processes \ref\studies{B. Grinstein and M.B. Wise, Phys. Lett. B201
(1988) 274\semi S. Bertolini, F. Borzumati and A. Masiero, Phys. Lett.
B192 (1987) 437\semi A.N. Amaglobeli, A.G. Liparteliani and G.G.
Volkov, Sov. J. Nucl. Phys. 49 (1989) 535\semi J. L. Hewett, Phys.
Lett. B193 (1987) 327}. Predictions of exclusive event rates are
uncertain, however, because they require the calculation of
nonperturbative hadronic matrix elements.  Inclusive rates, although
they may be calculated more reliably, are considerably more difficult
to measure.

In this paper, we will investigate the form factors which describe the
rare decay $\ol B\to \ol K\e^+\e^-$. Isgur and Wise \ref\iwone{N. Isgur
and M.B. Wise, Phys. Rev. D42 (1990) 2388} have used heavy quark spin
and flavor symmetries to relate the form factors for $\ol B\to \ol
K\e^+\e^-$ to those for semileptonic $D$ meson decay. Burdman and
Donoghue \ref\burd{G. Burdman and J.F. Donoghue, Phys. Lett. B270
(1991) 55} have instead related $\ol B\to \ol K\e^+\e^-$ to
semileptonic $\ol B$ meson decay. This approach may seem reasonable,
since it avoids the use of the heavy quark flavor symmetry, in
particular the question of whether the heavy quark limit is a good
approximation for charm. But the analysis invokes, as compensation,
chiral $SU(3)$ symmetry. It is the purpose of this letter to
investigate the validity of this latter approximation in this process.

We will compute violations to the $SU(3)$ symmetry limit, which arise
from the light quark masses $m_q$, by means of a phenomenological
lagrangian which displays simultaneously explicit chiral and heavy
quark symmetries. This lagrangian is non-renormalizable, and in order
to control the higher dimension terms, we consider only the portion of
the Dalitz plot in which the momentum of the kaon or pion is small. We
will compute one-loop expressions for the relevant form factors,
retaining only terms, such as those of the form $m_q\ln m_q$, which
depend nonanalytically on the symmetry breaking parameters. These terms
dominate the corrections in the theoretical limit of very small quark
masses, and they cannot be reabsorbed into counterterms at higher order
in the effective lagrangian.

With all these limitations, what is the interest in this computation?
Although the validity of the symmetry relations between $\ol B\to \ol
K\e^+\e^-$ and  $\ol B\to\pi\e^-\ol\nu$ form factors will not be fully
established, we will gain confidence in them if the nonanalytic
corrections are small. Alternatively, large (order 100\%) corrections
would be an immediate indication of the breakdown of the relations. In
this regard it is useful to keep in mind the case of the relation
between kaon decay and the parameter $B_K$, which is invalidated by
large corrections of precisely this sort \ref\bsw{J. Bijnens, H. Sonoda
and M.B. Wise, Phys. Rev. Lett. 53 (1984) 2367}.

The rare decay $\ol B\to \ol K\e^+\e^-$ occurs via the quark level
transitions $b\to s\,\gamma$ and $b\to s\,\e^+\e^-$.  These in turn are
induced by loop processes at the weak scale, appearing at low energies
as local nonrenormalizable operators with coefficients in which the
leading logarithms have been resummed \ref\gsw{B. Grinstein, M. Savage
and M.B. Wise, Nucl. Phys. B319 (1989) 271}. The three operators which
will be relevant here are
\eqn\oplist{\eqalign{
    \O_7&={e\over16\pi^2}m_b\,\ol\s_L\sigma^{\mu\nu}
    \b_R\,F_{\mu\nu}\,,\cr
    \O_8&={e^2\over16\pi^2}\,\ol\s_L\gamma^\mu\b_L\,
    \ol\e\gamma_\mu\e\,,\cr
    \O_9&={e^2\over16\pi^2}\,\ol\s_L\gamma^\mu\b_L\,
    \ol\e\gamma_\mu\gamma^5\e\,,\cr}}
assembled into an effective interaction Hamiltonian
\eqn\hamil{\H_{\rm eff}=-{4G_F\over\sqrt{2}}(s_3+s_2\e^{\im\delta})\,
    [c_7(\mu)\O_7+c_8(\mu)\O_8+c_9(\mu)\O_9]\,.}
The total rate for the decay $\ol B\to \ol K\e^+\e^-$ is calculated from
the matrix elements of these operators.  The part of the computation
which involves the leptons is perturbative and straightforward; however
the same may not be said for the matrix elements of the flavor-changing
quark operators between external hadron states.  These typically must
be parameterized in terms of a Lorentz-covariant decomposition,
\eqn\formfactors{\eqalign{
    \langle \ol K(p_K)\,|\,\ol\s\gamma^\mu\b\,|\,\ol B(p_B)\rangle
    &=f_+\,(p_B+p_K)^\mu + f_-\,(p_B-p_K)^\mu\,,\cr
    \langle \ol K(p_K)\,|\,\ol\s\sigma^{\mu\nu}\b\,|\,\ol B(p_B)\rangle
    &=\im h\,[(p_B+p_K)^\mu (p_B-p_K)^\nu -
    (p_B+p_K)^\nu (p_B-p_K)^\mu]\,,\cr}}
in which the form factors $f_+$, $f_-$ and $h$ are scalar functions of
the invariant momentum transfer $p_K\cdot p_B$.  The differential
partial decay width at fixed $\hat{s}=(p_{\e^+}+p_{\e^-})^2/m_B^2$ is
then given by
\eqn\diffrate{\eqalign{
    {\d\Gamma\over\d\hat{s}}(\ol B\to \ol K\e^+\e^-)=&\left| s_3+
    s_2\e^{\im\delta}\right|^2\,{G_F^2m_B^5\alpha^2\over3\cdot2^9\pi^5}
    \cr &\times\left[\left(1-m_K^2/m_B^2\right)^2-2\hat{s}
    \left(1+m_K^2/m_B^2\right)+\hat{s}^2\right]^{3/2}\cr
    &\times\left[\left|c_8(m_b)f_++2m_b c_7(m_b)h\right|^2 +
    \left| c_9(m_b)f_+\right|^2\right]\,.\cr}}
The coefficients $c_7(m_b)$, $c_8(m_b)$ and $c_9(m_b)$ depend on
short-distance physics and are discussed in detail in ref.~\gsw.

The form factors $f_+$ and $h$ which are needed for eq.~\diffrate\
involve nonperturbative strong interactions and are in general
incalculable.  However the fact that the bottom quark is very massive
compared to scales typical of QCD affords some simplifications,
\eqn\hqetrels{\eqalign{
    &f_++f_-\sim m_b^{-1/2},\qquad f_+-f_-\sim m_b^{1/2},\cr
    &h={f_+-f_-\over2m_b}\,, \qquad s=(f_++f_-)m_b+(f_+-f_-)p_K\cdot
    p_B/m_b\,.\cr}}
For completeness, we have included the scalar form factor $s$, which
parameterizes the matrix element $\langle \ol
K(p_K)\,|\,\ol\s\,\b\,|\,\ol B(p_B)\rangle$. Hence, in the simultaneous
limits of chiral symmetry and $m_b\to\infty$, the form factors for the
decay $\ol B\to \ol K\e^+\e^-$ are given simply in terms of the form
factor $f_+$ which describes $\ol B\to\pi\e^-\ol\nu$.

If we now restrict ourselves to that portion of the Dalitz plot in
which the leptons are emitted back to back, and the kaon is very soft,
we will be able to compute the hadronic matrix elements \formfactors\
in terms of two phenomenological parameters. These are the decay
constant $f_B$ of the $\ol B$ meson, and the axial coupling $g$ of the
pion to the $(\ol B$,$\ol B^*)$ doublet.  These constants appear as
coefficients in a nonrenormalizable low-energy effective lagrangian in
which both heavy quark and chiral $SU(3)$ symmetry are explicit.  This
is a framework within which the relations \hqetrels\ arise naturally,
and which also will allow us to compute the leading nonanalytic
corrections which test the validity of $SU(3)$ symmetry in this process.

We begin with a brief synopsis of the formalism of heavy hadron chiral
perturbation theory \ref\wise{M.B. Wise, Phys. Rev. D45 (1992)
2188\semi G. Burdman and J.F. Donoghue, Phys. Lett. B280 (1992) 287\semi
T.M. Yan \etal, Phys. Rev. D46 (1992) 1148}.
In the limit $m_b\to\infty$, the $\ol B$ and the $\ol B^*$ mesons are
degenerate, and to implement the heavy quark symmetries it is
convenient to assemble them into a ``superfield'' $H_a(v)$:
\eqn\hadef{H_a(v)={1+\vslash\over2\sqrt2}\left[
    \ol B_a^{*\mu}\gamma_\mu-\ol B_a\gamma^5\right]\,.}
Here $v^\mu$ is the fixed four-velocity of the  heavy meson, and $a$
is a flavor $SU(3)$ index corresponding to the light antiquark. Because
we have absorbed mass factors $\sqrt{2m_B}$ into the fields, they have
dimension 3/2; to recover the correct relativistic normalization, we
will multiply amplitudes by $\sqrt{2m_B}$ for each external $\ol B$ or
$\ol B^*$ meson.

The chiral lagrangian contains both heavy meson superfields and
pseudogoldstone bosons, coupled together in an $SU(3)_L\times SU(3)_R$
invariant way.  The matrix of pseudogoldstone bosons appears in the
usual exponentiated form $\xi=\exp(\im{\cal M}/f)$, where
\eqn\cm{
    {\cal M}=\pmatrix{{\textstyle{1\over\sqrt{2}}}\pi^0+{\textstyle
    {1\over\sqrt{6}}}\eta&\pi^+&K^+\cr\pi^-
    &{\textstyle{-{1\over\sqrt{2}}}}\pi_0+
    {\textstyle{1\over\sqrt{6}}}\eta&K^0\cr K^-&\overline K^0&-
    {\textstyle\sqrt{2\over 3}}\,\eta}\,,}
and $f$ is the pion (or kaon) decay constant.  The bosons couple to the
heavy fields through the covariant derivative and axial vector field,
\eqn\covax{
    \eqalign{&D_{ab}^\mu=\delta_{ab}\del^\mu+V_{ab}^\mu
    =\delta_{ab}\del^\mu+\half\left(\xi^\dagger
    \del^\mu\xi+\xi\del^\mu\xi^\dagger\right)_{ab}\,,\cr
    &A_{ab}^\mu={\im\over 2}\left(\xi^\dagger\del^\mu\xi
    -\xi\del^\mu\xi^\dagger\right)_{ab}
    =-{1\over f}\partial_\mu{\cal M}_{ab}+{\cal O}({\cal M}^3)\,.}}
Lower case roman indices correspond to flavor $SU(3)$. Under chiral
$SU(3)_L\times SU(3)_R$, the pseudogoldstone bosons and heavy meson
fields transform as $\xi\to L\xi U^\dagger=U\xi R^\dagger$, $A^\mu\to
UA^\mu U^\dagger$, $H\to HU^\dagger$ and $(D^\mu H)\to (D^\mu
H)U^\dagger$, where the matrix $U_{ab}$ is a nonlinear function of the
pseudogoldstone boson matrix ${\cal M}$.

The chiral lagrangian is an expansion in derivatives and pion fields,
as well as in inverse powers of the heavy quark mass.  The kinetic
energy terms take the form
\eqn\original{
    \eqalign{{\cal L}_{\rm kin}&={1\over8}
    f^2\,\partial^\mu\Sigma_{ab}\,\partial_\mu\Sigma^{\dagger}_{ba}
    -\Tr\left[\overline H_a(v)\im v\cdot D_{ba} H_b(v)\right]\,,}}
where $\Sigma=\xi^2$.  The leading interaction term is of
dimension four,
\eqn\gdef{
    g\,\Tr\left[\overline H_a(v)H_b(v)\,\Aslash_{ba}\gamma^5\right]\,,}
where $g$ is an unknown parameter, of order one in the constituent
quark model. The analogue of this term in the charm system is
responsible for the decay $D^*\to D\pi$, from which one may derive the
limit $g^2<0.5$.

The quark bilinears $J^\mu=\ol\s\gamma^\mu\b$ and
$J^{\mu\nu}=\ol\s\sigma^{\mu\nu}\b$, whose hadronic matrix elements we
must compute, may be matched onto operators in the chiral lagrangian
written in terms of the meson fields.  Heavy quark symmetry and the
$SU(3)_L\times SU(3)_R$ transformation properties of chiral currents
dictate that this matching must to leading order take the universal form
\eqn\matchforms{\eqalign{
   \ol\q_{aL}\Gamma\b&\to c_L\,\Tr[\Gamma H_b(v)\xi^\dagger_{ba}]\,,\cr
   \ol\q_{aR}\Gamma\b&\to c_R\,\Tr[\Gamma H_b(v)\xi_{ba}]\,,\cr}}
for left- and right-handed light quark fields, where $\Gamma$ is an
arbitrary Dirac matrix.  Then the two conditions
\eqn\matchconds{\eqalign{
    \langle 0\,|\,\ol\q_a\gamma^\mu\gamma^5\b\,|\,\ol B_a(p)\rangle
    &=\im f_Bp^\mu\,,\cr
    \langle\pi(p')\,|\,\ol\q_a\gamma^\mu\gamma^5\b\,|\,\ol B_a(p)
    \rangle &=0\,,\cr}}
are sufficient to determine $c_L$ and $c_R$,
\eqn\clcr{c_L=c_R={\im\over2}f_B\sqrt{m_B}\,.}
As we are working in the $SU(3)$ limit, the decay constant $f_B$ is
flavor symmetric.  Note that the first of the conditions \matchconds\
is merely the definition of $f_B$, while the second reflects the
invariance under parity of the strong interactions.

Decomposing the bilinears $J^\mu$ and $J^{\mu\nu}$ into chiral
components, it is straightforward to perform the matching onto
interactions in the effective lagrangian.  We find the operators
\eqn\matching{\eqalign{
    \O^\mu &={\im\over4}f_B\sqrt{m_B}\,\left\{\Tr\left[\gamma^\mu
    H_b(v)(\xi^\dagger+\xi)_{ba}\right]+\Tr\left[\gamma^5\gamma^\mu
    H_b(v)(\xi^\dagger-\xi)_{ba}\right]\right\}\cr
    \O^{\mu\nu} &={\im\over4}f_B\sqrt{m_B}\,
    \left\{\Tr\left[\sigma^{\mu\nu}H_b(v)(\xi^\dagger+\xi)_{ba}\right]
    +\Tr\left[\gamma^5\sigma^{\mu\nu}H_b(v)(\xi^\dagger-\xi)_{ba}\right]
    \right\} \,.\cr}}
For the operators $J^\mu$ and $J^{\mu\nu}$, which carry strangeness, we
take $a=3$. Each of these relations is corrected at higher order in the
chiral derivative expansion.  Note that the first terms in \matching\
yield vertices with an even number of pseudogoldstone bosons, while the
second terms yield those with an odd number.

We are now in a position to compute the hadronic matrix elements
\formfactors\ in the effective theory.\foot{A recent preprint
\ref\casal{R. Casalbuoni \etal, UGVA-DPT 1993/04-816 and BARI-TH/93-140
(1993)} has come to our attention, which also considers the process $\ol
B\to \ol K\e^+\e^-$ in this theory.  However the authors compute
only the contributions of the operator $\O_7$, and they do not address
the issue of $SU(3)$ violating corrections.}  The tree level Feynman
diagrams are shown in \fig\tree{Tree level amplitudes for $\ol B\to
\ol K$.  The solid line represents the heavy meson, the dashed lines
pseudogoldstone bosons.  The solid square indicates the insertion of
the flavor-changing operator $\O^\mu$ or $\O^{\mu\nu}$.  (a) the pole
amplitude $\A_{\rm pole}$; (b) the point amplitude $\A_{\rm point}$.}.
There exist both pole graphs, \tree (a), in which a kaon is emitted
via the interaction \gdef\ and the virtual $\ol B_s^*$ meson is
absorbed by one of the effective operators, and direct graphs, \tree
(b), in which the effective operator both absorbs the $\ol B$ and
emits the $\ol K$. The former are induced by the first terms in
eqs.~\matching, while the latter are induced by the second terms.

It is extremely straightforward to compute the desired amplitudes.  For
the vector and tensor currents, respectively, we find for the pole
graphs
\eqn\poleamps{\eqalign{
    \A^\mu_{\rm pole}&=-{gf_Bm_B\over f}{1\over p_K\cdot v+\Delta}
    \,(p_K^\mu-p_K\cdot vv^\mu)\,,\cr
    \A^{\mu\nu}_{\rm pole}&=-{gf_Bm_B\over f}{1\over p_K\cdot v+\Delta}
    \,\im\,(p_K^\mu v^\nu-p_K^\nu v^\mu)\,,\cr}}
where $\Delta=m_{B_s^*}-m_B$, and $p_K\cdot v$ is the kaon energy in the
$\ol B$ rest frame.  For the point amplitudes, we find
\eqn\pointamps{\eqalign{
    \A^\mu_{\rm point}&=-{f_Bm_B\over f}v^\mu\,,\cr
    \A^{\mu\nu}_{\rm point}&=0\,.\cr}}
We may now solve for the form factors $f_+$, $f_-$ and $h$, obtaining
\eqn\formfacans{\eqalign{
    f_\pm&=-{f_B\over2f}\left[ 1\pm g{m_B\mp p_K\cdot v\over
    \Delta+p_K\cdot v}\right]\,,\cr
    h&=-{f_B\over2f}{g\over\Delta+p_K\cdot v}\,.}}
Note that in the form factors $f_\pm$, the pole amplitudes dominate the
direct ones by a factor $m_B/(p_K\cdot v+\Delta)$, so $f_\pm\to\mp
gf_Bm_B/2f(\Delta+p_K\cdot v)$ as $m_b\to\infty$. Substituting $f_+$
and $h$ into eq.~\diffrate, we may now compute the partial decay rate.
It is convenient to normalize to the semileptonic width $\Gamma(\ol
B\to X_c\e^-\ol\nu)$, after which we obtain
\eqn\diffanswer{\eqalign{
    {1\over\Gamma(\ol B\to X_c\e^-\ol\nu)}
    {\d\Gamma\over\d\hat{s}}(\ol B \to \ol K\e^+\e^-)
    =&{\alpha^2\over8\pi^2}
    {g^2f_B^2\over4f^2}{1\over(\hat{E}_K+\hat\Delta)^2}\cr
    &\times\left[\left(1-\hat{m}_K^2\right)^2-2\hat{s}
    \left(1+\hat{m}_K^2\right)+\hat{s}^2\right]^{3/2}\cr
    &\times\left[\left|c_8(m_b)\hat{f}_+(\hat{s})
    +2c_7(m_b)\right|^2+\left| c_9(m_b)\hat{f}_+(\hat{s})\right|^2
    \right]\,,\cr}}
where $\hat{m}_K^2=m_K^2/m_B^2$, $\hat\Delta=\Delta/m_B$ and
\eqn\eandf{\eqalign{
    \hat{E}_K&=E_K/m_B=(1+\hat{m}_K^2-\hat{s})/2\,,\cr
    \hat{f}_+&=1-\hat{E}_K+(\hat{E}_K+\hat\Delta)/g\,.\cr}}

Our results so far assume an exact $SU(3)$ chiral symmetry among the
light quarks.  The virtue of this effective lagrangian formalism is
that it allows us to make some estimate of the size of $SU(3)$
violating corrections.  Of course, the leading corrections typically
involve new terms in the chiral lagrangian, whose coefficients must be
fixed. Unfortunately, the current paucity of data on heavy meson
interactions with pseudogoldstone bosons precludes any experimental
determination of these coefficients.  However, there are certain
nonanalytic corrections, such as those of the form $m_\pi^2\ln
m_\pi^2$, which are independent of such new terms. These corrections
are determined uniquely by loops in the flavor-conserving effective
lagrangian, in which the $SU(3)$ violation enters indirectly via the
pseudogoldstone boson masses. While such chiral logarithms are in fact
dominant in the limit of very small light quark masses, for the
physical pions and kaons this is unlikely to be the case. Still, we may
hope that such loops at least indicate the magnitude of $SU(3)$
violation, even if they do not provide us with precise quantitative
information. In particular, if the nonanalytic corrections are large
($\sim100\%$), we will certainly know not to trust the results
\formfacans\ and the extrapolation of matrix elements from $\ol
B\to\pi$ to $\ol B\to \ol K$.  However, if they are small we may gain
some additional confidence that what we have done is sensible. In any
case, this is the spirit in which we shall proceed.

Since we expect the largest corrections to come from the large $K$ and
$\eta$ masses, it is appropriate to simplify the calculation by making
two approximations.  First, we shall set $m_{\pi^\pm}=m_{\pi^0}\equiv
m_\pi$ and $m_{K^\pm}=m_{K^0,\ol K^0}\equiv m_K$.  Second, we shall set
all mass splittings between the various flavor and spin states of the
$B$ mesons to zero when they appear in loops.  (Note that we do {\it
not\/} ignore the splitting $\Delta$ when it appears in a pole, as in
eq.~\poleamps.)  In order to focus on $SU(3)$ violation, we will
compute separately the corrections to the matrix elements for
$B^-\to\pi^-$ and $B^-\to K^-$. For each nonvanishing graph, we will
present the nonanalytic dependence on the pion masses and on the
momentum of the external pion or kaon, giving the answer as a
fractional correction to the tree level result.  At the end we will
assemble the various pieces and provide a numerical estimate of the
size of these leading nonanalytic contributions to the violation of
chiral $SU(3)$ symmetry.

It will be convenient to express the results in terms of a few general
Feynman integrals.  After applying dimensional regularization to the
ultraviolet divergences, there will be nonanalytic dependence not only
on the pion masses and the external momenta, but on the renormalization
scale $\mu$ as well.  Since it is precisely this behavior in which we
are interested, we will drop any additional constants which may appear.

The first two integrals have no Lorentz dependence.  They are
\eqn\ioneitwo{\eqalign{
    &\im\pintegral {1\over p^2-m^2}={1\over16\pi^2}I_1(m)
    +\ldots\,,\cr
    &\im\pintegral {1\over (p^2-m^2)(p\cdot v-\Delta)}=
    {1\over16\pi^2}{1\over\Delta}I_2(m,\Delta)+\ldots\,,\cr}}
where
\eqn\idefs{\eqalign{
    I_1(m)&=m^2\ln(m^2/\mu^2)\,,\cr
    I_2(m,\Delta)&=-2\Delta^2\ln(m^2/\mu^2)-4\Delta^2F(m/\Delta)\,.\cr}}
The function $F(x)$ will appear frequently.  It is most convenient to
write it in a form where the smooth transition between the regimes
$x<1$ and $x>1$ is apparent:
\eqn\fdef{F(x)=\left\{\eqalign{
    &\sqrt{1-x^2}\;\tanh^{-1}\sqrt{1-x^2}\,,\cr
    &-\sqrt{x^2-1}\;\tan^{-1}\sqrt{x^2-1}\,,\cr}
    \qquad\eqalign{&x\le1\cr &x\ge1\cr}\right.}
The third integral is a two-index symmetric tensor:
\eqn\jmunu{\eqalign{
    J^{\mu\nu}(m,\Delta)&=\im\pintegral{p^\mu p^\nu\over(p^2-m^2)
    (p\cdot v-\Delta)}\cr
    &={1\over16\pi^2}\Delta\left[J_1(m,\Delta)g^{\mu\nu}
    +J_2(m,\Delta)v^\mu v^\nu\right]+\ldots\,,\cr}}
where
\eqn\joneandtwo{\eqalign{
    J_1(m,\Delta)&=(-m^2+\frac23\Delta^2)\ln(m^2/\mu^2)+\frac43
    (\Delta^2-m^2)F(m/\Delta)\,,\cr
    J_2(m,\Delta)&=(2m^2-\frac83\Delta^2)\ln(m^2/\mu^2)-\frac43
    (4\Delta^2-m^2)F(m/\Delta)\,.\cr}}
Finally, we have an integral which can be derived from $J^{\mu\nu}$,
\eqn\kmunu{\eqalign{
    K^{\mu\nu}(m,\Delta_1,\Delta_2)&=\im\pintegral{p^\mu p^\nu\over
    (p^2-m^2)(p\cdot v-\Delta_1)(p\cdot v-\Delta_2)}\cr
    &={1\over16\pi^2}\left[K_1(m,\Delta_1,\Delta_2)g^{\mu\nu}
    +K_2(m,\Delta_1,\Delta_2)v^\mu v^\nu\right]+\ldots\cr
    &={1\over\Delta_1-\Delta_2}\left[J^{\mu\nu}(m,\Delta_1)
    -J^{\mu\nu}(m,\Delta_2)\right]\,.\cr}}
We will need only the limit
$K(m,\Delta)=K_1(m,\Delta,0)$, which takes the simple form
\eqn\kdef{K(m,\Delta)=J_1(m,\Delta)-{2\pi\over3}
    {m^3\over\Delta}\,,}
and we note that $K(m,0)=-I_1(m)$.

With these integrals in hand, we now turn to the set of Feynman graphs
which we must compute.  The diagrams fall into three classes:  those
which correct the pole amplitudes $\A_{\rm pole}$, those which correct
the point amplitudes $\A_{\rm point}$, and those which correct both.
In the last class is the wavefunction renormalization of the
$B^-$ meson, depicted in \fig\wavefun{Diagrams contributing to the
wavefunction renormalization of the external $B^-$.  (a) correction to
$\A_{\rm pole}$; (b) correction to $\A_{\rm point}$.}. This graph is
universal, independent of the external pion momentum or flavor.  The
result may be obtained from ref.~\ref\fbcalc{B. Grinstein \etal, Nucl.
Phys. B380 (1992) 369\semi P. Cho, Harvard University preprint
HUTP-92-A039 (1992)\semi A.F. Falk, Phys. Lett. B305 (1993) 268}.  For
both $\A_{\rm pole}$ and $\A_{\rm point}$, we find a fractional
correction to the tree amplitude of
\eqn\wavegraph{
    \gpref\left[-\frac94I_1(m_\pi)-\frac32I_1(m_K)-\frac14I_1(m_\eta)
    \right]\,.}

There are two nonzero graphs which correct the point amplitude $\A_{\rm
point}$, depicted in \fig\pointcorrs{Diagrams which correct the point
amplitude $\A_{\rm point}$.}. Although we have seen that the form
factors of interest are actually dominated by the pole amplitude, we
will include these diagrams for completeness. The diagram in
\pointcorrs (a) yields a fractional correction to the matrix element
for $B^-\to\pi^-$ of
\eqn\threeapion{
    \pref \left[-\frac5{12}I_1(m_\pi)-\frac13I_1(m_K)
    -\frac1{12}I_1(m_\eta)\right]\,,}
while for $B^-\to K^-$ the result is
\eqn\threeakaon{
    \pref \left[-\frac14I_1(m_\pi)-\frac12I_1(m_K)
    -\frac1{12}I_1(m_\eta)\right]\,.}
The graph in \pointcorrs (b) requires a two-pion interaction which
arises from the $V^\mu_{ab}$ part of the heavy meson kinetic energy
term \original.  It also depends on $E_\pi=p_\pi\cdot v$, the energy of
the external pion (or kaon) in the rest frame of the $B^-$. For
$B^-\to\pi^-$, we find
\eqn\threebpion{
    \pref\left[I_1(m_\pi)+\frac12I_1(m_K)
    +2I_2(m_\pi,E_\pi)+I_2(m_K,E_\pi)\right]\,.}
For $B^-\to K^-$ we obtain
\eqn\threebkaon{
    \pref\left[I_1(m_K)+\frac12I_1(m_\eta)
    +2I_2(m_K,E_K)+I_2(m_\eta,E_K)\right]\,.}
The diagrams in \pointcorrs (c) and \pointcorrs (d) vanish.

There are four nonzero graphs which correct the pole amplitude
$\A_{\rm pole}$, depicted in \fig\polecorrs{Diagrams which correct the
pole amplitude $\A_{\rm pole}$.}. The diagram in \polecorrs (a) is
simple, since it is independent of the external pion momentum.  For
$B^-\to\pi^-$ we find the fractional correction
\eqn\fourapion{
    \pref\left[-\frac34I_1(m_\pi)-\frac12I_1(m_K)
    -\frac1{12}I_1(m_\eta)\right]\,,}
while for $B^-\to K^-$ we obtain
\eqn\fourakaon{
    \pref\left[-I_1(m_K)-\frac13I_1(m_\eta)\right]\,.}
The graph in \polecorrs (b) is equally straightforward.  The correction
to $B^-\to\pi^-$ is given by
\eqn\fourbpion{
    \pref\left[-\frac23I_1(m_\pi)-\frac13I_1(m_K)\right]\,,}
while for $B^-\to K^-$ it is
\eqn\fourbkaon{
    \pref\left[-\frac14I_1(m_\pi)-\frac12I_1(m_K)
    -\frac14I_1(m_\eta)\right]\,.}
The diagrams in \polecorrs (c) and (d) actually consist of two graphs.
Since the interaction term \gdef\ contains a $\ol B^*$-$\ol B^*$-$\pi$
coupling as well as $\ol B^*$-$\ol B$-$\pi$, the heavy meson line can
take either the form $\ol B$---$\ol B^*$---$\ol B$---$\ol B^*$ or the
form $\ol B$---$\ol B^*$---$\ol B^*$---$\ol B^*$. In \polecorrs (c) the
second possibility gives twice the former.  We find a somewhat more
complicated dependence on the external momentum $p_\pi\cdot v$, which
is expressed in terms of the integral $J^{\mu\nu}$.  However, we can
resum this contribution into the denominator of the $\ol B^*$
propagator, at which point it is consistent with our approximations to
subtract the term which renormalizes the meson mass.  This procedure
introduces the limit $K(m,\Delta)$ of the general integral
$K^{\mu\nu}$. For $B^-\to\pi^-$ we then find the correction
\eqn\fourcpion{
    \gpref\left[\frac92K(m_\pi,E_\pi)+3K(m_K,E_\pi)
    +\frac12K(m_\eta,E_\pi)\right]\,,}
and for $B^-\to K^-$ we obtain
\eqn\fourckaon{
    \gpref\left[6K(m_K,E_K)+2K(m_\eta,E_K)\right]\,.}
In \polecorrs (d), the second possibility gives minus twice the first.
The momentum dependence enters through the limit $K(m,\Delta)$ of the
general integral $K^{\mu\nu}$. The fractional correction to
$B^-\to\pi^-$ may then be written
\eqn\fourdpion{
    \gpref\left[-\frac12K(m_\pi,E_\pi)
    +\frac16K(m_\eta,E_\pi)\right]\,,}
while for $B^-\to K^-$ we obtain
\eqn\fourdkaon{\gpref\left[-\frac13K(m_\eta,E_K)\right]\,.}
The diagrams in \polecorrs (e)--(g) vanish identically.

Finally, for both $\A_{\rm pole}$ and $\A_{\rm point}$ we must include
the wavefunction renormalization of the external pseudogoldstone boson,
as shown in \fig\pgbwavef{Diagrams contributing to the wavefunction
renormalization of the external pseudogoldstone boson.  (a) correction
to $\A_{\rm pole}$; (b) correction to $\A_{\rm point}$.}.  The pion
self-interaction is induced by the kinetic energy term \original.  For
$B^-\to\pi^-$ we find the fractional correction
\eqn\fivepion{
    \pref\left[-\frac23I_1(m_\pi)-\frac13I_1(m_K)\right]\,,}
while for $B^-\to K^-$ we obtain
\eqn\fivekaon{
    \pref\left[-\frac14I_1(m_\pi)-\frac12I_1(m_K)-\frac14I_1(m_\eta)
    \right]\,.}

We now assemble these various amplitudes into an estimate of the size
of $SU(3)$ corrections in this process.  We begin with the pole
amplitudes, because they dominate the observable form factors in the
limit $m_b\to\infty$.  Although one could simply add together the
diagrams in \wavefun, \polecorrs\ and \pgbwavef, it is more reasonable
to absorb some of the corrections into a renormalization of the heavy
meson decay constant $f_B$.  Since in $\A_{\rm pole}$ the pion or kaon
is emitted before the flavor-changing operator $\O^\mu$ or
$\O^{\mu\nu}$ acts, it is either $f_{B_d}$ (for $B^-\to\pi^-$) or
$f_{B_s}$ (for $B^-\to K^-$) which is relevant to the amplitude.  In
fact, this would be precisely the combined effect of \polecorrs (a) and
half of \polecorrs (c), if the momentum of the external pion or kaon
were set to zero. The relation between the bare parameter $f_B$ and the
renormalized decay constants, computed in the same chiral logarithmic
approximation, is given by \fbcalc
\eqn\fbrels{\eqalign{
    f_B&=f_{B_d}\left\{1+\pref\left(\frac12+\frac32g^2\right)
    \left[\frac32I_1(m_\pi)+I_1(m_K)+\frac16I_1(m_\eta)\right]
    \right\}\,,\cr
    f_B&=f_{B_s}\left\{1+\pref\left(\frac12+\frac32g^2\right)
    \left[2I_1(m_K)+\frac23I_1(m_\eta)\right]\right\}\,.\cr}}
Similarly, it is appropriate to renormalize the pseudogoldstone boson
decay constant $f$ to $f_\pi$ or $f_K$, for which we have \ref\fpi{P.
Langacker and H. Pagels, Phys. Rev. D8 (1973) 4595}
\eqn\frels{\eqalign{
    f&=f_\pi\left\{1-\pref\left[2I_1(m_\pi)+I_1(m_K)
    \right]\right\}\,,\cr
    f&=f_K\left\{1-\pref\left[\frac34I_1(m_\pi)+\frac32I_1(m_K)
    +\frac34I_1(m_\eta)\right]\right\}\,.\cr}}
Note that in the amplitudes $\A_{\rm pole}$ and $\A_{\rm point}$, $f$
appears in the denominator.

In estimating the diagrams, we take the masses $m_\pi=140\mev$,
$m_K=490\mev$, and $m_\eta=550\mev$.  Since the largest corrections are
come from the $K$ and $\eta$ masses, we take the pseudogoldstone boson
decay constant $f$ to be $f_K\approx165\mev$.  To be conservative, we
take the renormalization scale $\mu=1\gev$, since this choice magnifies
the effect of the chiral logarithms.  For the same reason we choose the
coupling $g$ to be as large as possible; since from the width for
$D^*\to D\pi$ we have $g^2\le0.5$, we take $g^2=0.5$ in our
estimates.\foot{In fact, we would be justified in using the amplitudes
in \wavefun\ and \polecorrs (b)--(d) to correct the prediction for
$D^*\to D\pi$ and extract experimentally a ``renormalized'' $g$.  Doing
this would tighten the experimental upper limit on $g$ by approximately
$15\%$; instead of the limit $g^2<0.5$, we would have $g^2<0.4$.
However, to be conservative, we do not include this additional
restriction here.} Finally, when they appear we take the external
pseudogoldstone boson energies to be equal to their masses,
$E_\pi=m_\pi$.  This is consistent with the soft pion limit in which we
are working, and simplifies our estimates.

Assembling the corrections as we have described, and replacing $f_B\to
f_{B_d}$, $f\to f_\pi$ in $B^-\to\pi^-$ and $f_B\to f_{B_s}$, $f\to
f_K$ in $B^-\to K^-$, we obtain a residual correction to the dominant
pole amplitudes $\A_{\rm pole}$, for $B^-\to\pi^-$ of $-13\%$ and for
$B^-\to K^-$ of $-51\%$.  Hence, in this approximation where we keep
only the nonanalytic dependence on the masses, we find $SU(3)$
violation at the level of $\sim40\%$. For the point amplitudes $\A_{\rm
point}$, we must include the diagrams in \wavefun, \pointcorrs\ and
\pgbwavef, plus the decay constant redefinitions~\fbrels\ and \frels.
We then find a correction of $1\%$ to the amplitude for $B^-\to\pi^-$,
while the correction to $B^-\to K^-$ is $13\%$.

Finally, we may use our results to estimate the $SU(3)$ corrections to
the coupling constant $g$ which multiplies the interaction term \gdef.
This is given by the graphs in \polecorrs (b) and (d), plus the
wavefunction renormalization on the external meson (\wavefun (a))
and pseudogoldstone boson (\pgbwavef (a)) lines.  The only new piece is
the $B_s$ wavefunction renormalization; like that for the $B^-$, it may
be obtained from ref.~\fbcalc, and is given by
\eqn\bswave{
    \gpref\left[-3I_1(m_K)-I_1(m_\eta)\right]\,.}
The tree level amplitude due to the interaction \gdef\ is proportional
to $g/f$; at one loop, for an external pion this will become
$g_\pi/f_\pi$, and for an external kaon $g_K/f_K$.  Hence we must also
include the correction \frels\ in computing $g_\pi$ and $g_K$.
Assembling the results, we find $g_\pi\approx1.14\,g$ and
$g_K\approx1.21\,g$.  This effect is in part $SU(3)$ symmetric; $SU(3)$
violation appears at the level of only $\sim7\%$.  As we have noted,
however, the $SU(3)$ conserving correction has an impact on the
extraction of the parameter $g$ from the decay $D^*\to D\pi$.

The violation of $SU(3)$ symmetry at the $40\%$ level which we have
found is substantial, but not necessarily so much so that we would
consider the entire computation to be untrustworthy.  Indeed, we do not
find nonanalytic corrections at the level of $100\%$, such as plague
other processes. Of the $40\%$ correction, half of it comes from
resolving the flavor ambiguities in the decay constants via the
replacements \fbrels\ and \frels.  Of course, we should stress that by
itself the computation of the nonanalytic corrections proves nothing,
since the analytic corrections due to higher order terms in the
phenomenological lagrangian could still be large and spoil the desired
relations.  Rather, we view our calculation as helping to build
confidence that using $SU(3)$ symmetry to compute the form factors for
$\ol B \to \ol K\e^+\e^-$ may indeed be a sensible treatment of the
nonperturbative matrix elements.

\bigskip
A.~F.~and B.~G.~acknowledge the support of the Department of Energy,
under contracts DE--AC03--76SF00515 and DE--AC35--89ER40486,
respectively.

\listrefs
\listfigs
\vfil\eject
\bye